\documentstyle[preprint,aps]{revtex}	% PREPRINT STYLE
\begin{document}
\draft
%
%%%%%%%%%%%%%%%%%%%%% TITLE PAGE %%%%%%%%%%%%%%%%%%%%%%%%%%%%%%%%%%%%%%%%%%%%%%
%
\preprint{UCD--94--29}
%\preprint{\today}
%
\title{Hadronic $Z\gamma$ Production with\\
QCD Corrections and Leptonic Decays}
\author{J.~Ohnemus}
\address{
Department of Physics\\
University of California\\
Davis, CA 95616 USA}
\maketitle
\begin{abstract}
The process $p\,p\hskip-7pt\hbox{$^{^{(\!-\!)}}$} \rightarrow  Z
\gamma + X \rightarrow \ell^- \ell^+ \gamma + X$, where $\ell$ denotes
a lepton,  is calculated to ${\cal O}(\alpha_s)$. Total and
differential cross sections, with  acceptance cuts imposed on the
leptons and photon, are given for the Tevatron and LHC center of mass
energies. In general, invariant mass and angular distributions are
simply scaled up in magnitude by the QCD radiative corrections,
whereas in transverse momentum distributions, the QCD radiative
corrections increase with the transverse momentum.
\end{abstract}
\vskip .5in
\newpage
%
%%%%%%%%%%%%%%%%%% MAIN TEXT %%%%%%%%%%%%%%%%%%%%%%%%%%%%%%%%%%%%%%%%%%%%%%%
%
\narrowtext

\section{INTRODUCTION}

Measurements of weak boson pair production at hadron colliders are
vital for testing the standard model and probing beyond it.  These
processes are particularly important because they can be used to probe
the  electroweak symmetry breaking mechanism and test the  triple weak
boson couplings \cite{EHLQ}. The process
$p\,p\hskip-7pt\hbox{$^{^{(\!-\!)}}$} \rightarrow Z \gamma$  is of
interest as a test of the standard model. In addition, this process
is sensitive to contact interactions which appear in composite models
of gauge bosons \cite{LEURER}, and it can also be used to search for
evidence of anomalous $ZZ\gamma$ and $Z\gamma\gamma$ couplings
\cite{ANOMALOUS}. In order to perform such tests it is imperative to
have precise calculations of hadronic $Z\gamma$ production.
Futhermore, the calculations should include the leptonic decay of the
$Z$ boson since the $Z$ boson is identified by its leptonic decay
products.

At leading order in QCD [${\cal O}(\alpha_s^0)$],  hadronic $Z\gamma$
production proceeds via the quark-antiquark annihilation subprocess $q
\bar q \to Z\gamma$.  This process was first calculated in
Ref.~\cite{RENARD}. $Z\gamma$ events can also be produced at leading
order by the photon bremsstrahlung process \cite{BREM} which proceeds
via subprocesses such as  $qg \to Zq$ followed by photon
bremsstrahlung from the final state quark. This process becomes
significant only at supercollider center of mass energies. The
production of $Z\gamma$ in association with one or two jets was
calculated in Refs.~\cite{VVONEJET} and  \cite{VVTWOJET},
respectively.   The gluon fusion subprocess $gg \to Z\gamma$, which
proceeds via a quark box loop, was calculated in
Refs.~\cite{AGPT,GLUONFUSION}.   Although this process is of order
$\alpha_s^2$, it can be important at supercollider energies due to the
large gluon luminosity;  the gluon fusion cross section is 15--30\% as
large as the $q\bar q \to Z\gamma$ Born cross section at supercollider
energies \cite{GLUONFUSION}.   The QCD radiative corrections to
hadronic $Z\gamma$ production were calculated in Ref.~\cite{ZGAMMA}
for the case of a real $Z$ boson in the final state.

In this paper, the calculation in Ref.~\cite{ZGAMMA}  is extended to
include the leptonic decay of the $Z$ boson.  Since the $Z$ boson is
identified via its leptonic decay products, the inclusion of the
leptonic decay in the calculation makes it much more useful for
comparing with experimental data.  For example, cuts can now be
applied to the final state leptons, thus allowing one to more closely
mimic the actual experimental conditions.

To date, the CDF and D0 experiments at the Fermilab Tevatron collider
have collected 12 \cite{ERREDE} and 6 \cite{DIEHL} $Z\gamma$ events,
respectively. Both experiments are expected to collect data
corresponding to an integrated luminosity of approximately
100~pb$^{-1}$ by the end of the current collider run (Run 1--B). This
represents a five fold increase in statistics over the presently
available data sample.  Larger cross sections and luminosities are
expected at the LHC. Thus it should soon be possible to use hadronic
$Z\gamma$ production to probe the standard model.

The remainder of this paper is organized as follows.  The formalism
used in the calculation is briefly reviewed in Sec.~II,  numerical
results for the Tevatron and LHC center of mass energies are given in
Sec.~III, and summary remarks are given in Sec.~IV.

\section{FORMALISM}

The formalism used in the calculation is reviewed in this section. The
discussion will be brief since the formalism has already been
described in the literature. The calculation is done by using a Monte
Carlo method for next-to-leading-order (NLO) calculations \cite{NLOMC}
in combination with helicity amplitude methods \cite{HELICITY}.
Details on the NLO calculation of hadronic $Z\gamma$ production  for a
real $Z$ boson in the final state can be found in Ref.~\cite{ZGAMMA}
and helicity amplitude calculations of $Z\gamma$ production are
described in Ref.~\cite{VVTWOJET}.

The calculation is done using the narrow width approximation for  the
decaying $Z$ boson. This simplifies the calculation greatly for two
reasons.  First of all, it is possible to ignore Feynman diagrams in
which the photon is radiated off one of the final state lepton lines
without violating electromagnetic gauge invariance.  [Radiative $Z$
decay events can be suppressed by a suitable choice of cuts; see
Sec.~IIIB.] Secondly, in the narrow width approximation it is
particularly easy to extend the NLO calculation of Ref.~\cite{ZGAMMA}
to include the leptonic decay of the $Z$ boson.

The QCD radiative corrections to hadronic $Z\gamma$ production were
calculated in Ref.~\cite{ZGAMMA} for the case of a real $Z$ boson in
the final state. The present calculation extends the results of
Ref.~\cite{ZGAMMA} to include the leptonic decay of the $Z$ boson. The
procedure for incorporating the leptonic decay $Z \to \ell^- \ell^+$
into the NLO calculation of hadronic $Z\gamma$ production is the same
as that used in Ref.~\cite{WGAMMANS} for incorporating the leptonic
decay $W \to \ell \nu$ into a NLO calculation of hadronic $W\gamma$
production. A detailed discussion of the procedure is given in
Ref.~\cite{WGAMMANS}. Basically, except for the virtual contribution,
all the NLO contributions for real $Z \gamma$ production have the form
\widetext
\begin{eqnarray}
d\sigma^{\hbox{\scriptsize NLO}}  (q \bar q \to Z \gamma) =
d\sigma^{\hbox{\scriptsize Born}} (q \bar q \to Z \gamma) \>
\left[ 1 + C_F \, {\alpha_s \over 2 \pi} (\ \ldots \ ) \right] \>,
\label{EQ:NLOFORM}
\end{eqnarray}
\narrowtext
where $\sigma^{\hbox{\scriptsize Born}}$ is the lowest order Born
cross section, $C_F = 4/3$ is the quark-gluon vertex color factor, and
$\alpha_s$ is the strong running coupling. Thus the leptonic decays
can be incorporated by simply making the replacement
\widetext
\begin{eqnarray}
d\sigma^{\hbox{\scriptsize Born}} (q \bar q \to Z \gamma) \longrightarrow
d\sigma^{\hbox{\scriptsize Born}} (q \bar q \to Z \gamma
\to \ell^- \ell^+ \gamma) \>
\label{EQ:NLOREPLACE}
\end{eqnarray}
\narrowtext
in the formulas of Ref.~\cite{ZGAMMA} for NLO real $Z \gamma$
production.

The simple replacement described in the previous paragraph does not
hold for the virtual correction. For this contribution we   use the
virtual correction for a real on-shell $Z$ boson which we subsequently
decay ignoring spin correlations. Since the virtual corrections are
small [about  1\% as large as the Born cross section]  and the effects
of spin correlations are small, the overall result of ignoring spin
correlations in the virtual correction is negligible compared to the
20\% -- 30\% uncertainty from the parton distribution functions and
the choice of the scale $Q^2$. [Note that spin correlations are
included everywhere in the calculation except in the virtual
contribution.]

\section{NUMERICAL RESULTS}

Numerical results for NLO $Z\gamma$ production at the Tevatron
[$p\,\bar p$ collisions at $\sqrt{s} = 1.8$~TeV] and the LHC [$p\,p$
collisions at $\sqrt{s} = 14$~TeV] are presented in this section. This
section begins with a brief description of the input parameters and
acceptance cuts.

\subsection{Input Parameters}

The numerical results presented in this section were obtained using
the two-loop expression for $\alpha_s$. The QCD scale
$\Lambda_{\hbox{\scriptsize QCD}}$  is specified for four flavors of
quarks by the choice of the parton distribution functions and is
adjusted whenever a heavy quark threshold is crossed so that
$\alpha_s$ is a continuous function of $Q^2$. The heavy quark masses
were taken to be $m_b=5$~GeV and $m_t=174$~GeV \cite{TOPQUARK}. The
standard model  parameters used in the numerical simulations are $M_Z
= 91.173$~GeV, $M_W = 80.22$~GeV, $\alpha (M_Z) =1/128$, and $\sin^2
\theta_{\hbox{\scriptsize w}} = 1 - (M_W^{}/M_Z^{})^2$. These values
are consistent with recent measurements at the CERN $e^+e^-$ collider
LEP \cite{LEP}, the Stanford Linear Collider \cite{SLD},  the CERN
$Sp\overline{p}S$ collider \cite{SPPS},  and the Fermilab Tevatron
collider \cite{TEVATRON}. The soft and collinear  cutoff parameters
used in the NLO Monte Carlo formalism  (see Ref.~\cite{ZGAMMA}) are
fixed to $\delta_s = 10^{-2}$ and $\delta_c =  10^{-3}$.  The parton
subprocesses have been summed over $u,d,s$, and $c$ quarks.  The
leptonic branching fractions and the total width of the $Z$ boson are
$B(Z \to e^- e^+) = B(Z \to \mu^- \mu^+) = 0.034$  and $\Gamma_Z =
2.487$~GeV, respectively.  A single scale $Q^2=M^2_{Z\gamma}$, where
$M_{Z\gamma}$ is the invariant mass of the $Z\gamma$ pair, has been
used for the renormalization scale $\mu^2$ and the factorization scale
$M^2$ (see Ref.~\cite{ZGAMMA} for a definition of these scales).  The
numerical results were obtained using the  Martin-Roberts-Stirling
(MRS)~\cite{MRSA} set A parton  distribution functions with $\Lambda_4
= 230$~MeV. These distribution functions have been fit to
next-to-leading order in the ${\rm \overline{MS}}$ (Modified Minimal
Subtraction) scheme \cite{MSBAR},  which is the scheme used for the
present calculation.

\subsection{Cuts}

The cuts imposed in the numerical simulations are motivated by  the
finite acceptance and resolution of the detector. The finite
acceptance of the detector is simulated by cuts on the four-vectors of
the final state particles.  These cuts include restrictions on the
transverse momentum $p_T^{}$  and rapidity $y$ of the photon and
leptons.

Since detectors are generally best equipped to measure photons which
are isolated,  {\it i.e.}, not accompanied by a large amount of
nearby energy \cite{ISOLATED1}, a photon isolation cut will be applied
in the numerical simulation. First, the leptons and photon are
required to be separated in the pseudorapidity-azimuthal-angle plane
\begin{eqnarray}
\noalign{\vskip 5 pt}
\Delta R (\ell,\gamma) =
  \left[ \left(\Delta \phi_{\ell \gamma}^{} \right)^2
+ \left(\Delta \eta_{\ell \gamma}^{} \right)^2
\right]^{1/2} \>.
\end{eqnarray}
A separation cut of $\Delta R(\ell,\gamma)>0.7$ will be imposed. In
addition,  a photon isolation cut typically  requires the sum of the
hadronic energy $E_{\hbox{\scriptsize had}}$ in a cone of size $R_0$
about the direction of the photon to be less than a fraction
$\epsilon_{\hbox{\scriptsize h}}$ of  the photon energy $E_{\gamma}$,
{\it i.e.},
\begin{eqnarray}
\sum_{\Delta R < R_0} \, E_{\hbox{\scriptsize had}} <
\epsilon_{\hbox{\scriptsize h}} \, E_{\gamma} \>,
\label{EQ:ISOL}
\end{eqnarray}
with $\Delta R = [(\Delta \phi)^2 + (\Delta \eta)^2 ]^{1/2}$.  A
photon isolation cut with $\epsilon_{\hbox{\scriptsize
h}}=0.15$~\cite{ISOLATED1,ISOLATED2}  and $R_0 = 0.7$ will be applied
in the numerical results presented in this section. This cut
suppresses the portion of the cross section that comes from the photon
bremsstrahlung process, which is beneficial since this process is not
well understood theoretically.

Before summarizing the cuts, it is useful to discuss cuts which will
suppress radiative $Z$ decays. Since photon radiation from the final
state lepton lines is ignored in the calculation, it is necessary to
impose cuts which will efficiently suppress  contributions from these
diagrams. In radiative $Z$ decays the lepton-photon separation in
$\Delta R$ space peaks sharply at small values due to the collinear
singularity associated with the diagrams in which the photon is
radiated from a final state lepton line.   A separation cut on $\Delta
R(\ell,\gamma)$, which has already been adopted, will  therefore
suppress radiative $Z$ decay events. Contributions from radiative $Z$
decay  can be further reduced by a cut on the $\ell^- \ell^+ \gamma$
invariant mass. In radiative $Z$ decays the $\ell^- \ell^+ \gamma$
system has  an invariant mass $M(\ell^- \ell^+ \gamma)$ close to
$M_Z^{}$,  whereas for $Z\gamma$ production $M(\ell^- \ell^+ \gamma)$
is always larger  than $M_Z^{}$ if finite $Z$-width effects are
ignored.  Thus $\ell^- \ell^+ \gamma$ events originating from
radiative $Z$ decays can be suppressed if $M(\ell^- \ell^+ \gamma)$ is
required to be slightly above $M_Z^{}$. Note that this requirement is
automatically satisfied in the present calculation since putting the
$Z$ boson on the mass shell and demanding the photon to have
$p_T^{}(\gamma) > p_T^{\hbox{\scriptsize min}}$  requires that
\begin{eqnarray}
M(\ell^- \ell^+ \gamma) > p_T^{\hbox{\scriptsize min}}
+ \left[ M_Z^2 + (p_T^{\hbox{\scriptsize min}})^2 \right]^{1/2} \>.
\end{eqnarray}
Thus no additional cuts are needed in the numerical simulation to
suppress the radiative $Z$ decays.

The complete set of cuts can now be summarized as follows.
\begin{quasitable}
\begin{tabular}{cc}
Tevatron & LHC\\
\tableline
$p_{T}^{}(\gamma) > 10$~GeV     & $p_{T}^{}(\gamma) >  50$~GeV\\
$p_{T}^{}(\ell)   > 20$~GeV     & $p_{T}^{}(\ell)   >  25$~GeV\\
$|y(\gamma)|      < 1.0$        & $|y(\gamma)|      <  2.5$\\
$|y(\ell)|        < 2.5$        & $|y(\ell)|        <  3.0$\\
$\Delta R (\ell,\gamma) > 0.7$  & $\Delta R (\ell,\gamma) > 0.7$\\
${\sum \atop \Delta R < 0.7} \, E_{\hbox{\scriptsize h}}
 < 0.15 \, E_{\gamma}$ &
${\sum \atop \Delta R < 0.7} \, E_{\hbox{\scriptsize h}}
 < 0.15 \, E_{\gamma}$
\end{tabular}
\end{quasitable}

\subsection{Cross Sections}

The total LO and NLO cross sections for  the process
$p\,p\hskip-7pt\hbox{$^{^{(\!-\!)}}$} \rightarrow  Z \gamma + X
\rightarrow \ell^- \ell^+ \gamma + X$ are given in Table~1 for center
of mass energies corresponding to the present Tevatron [$\sqrt{s} =
1.8$~TeV], an upgraded Tevatron [$\sqrt{s} = 3.5$~TeV], and the
proposed LHC [$\sqrt{s} = 14$~TeV]. The cross sections in Table~1 have
been summed over $\ell = e, \mu$ and include the cuts listed in
Sec.~IIIB. Note that  the LHC cross sections are smaller than the
Tevatron cross sections due to  the more restrictive $p_T^{}$ cuts
used for the LHC case [in particular the $p_T^{}(\gamma)$ cut].

The differential cross sections presented here have also been summed
over $\ell = e, \mu$ and  include the cuts listed in Sec.~IIIB.   The
figures are arranged in two parts, with parts a) and b) being  the
results for the Tevatron [$\sqrt{s} = 1.8$~TeV] and  LHC [$\sqrt{s} =
14$~TeV] energies, respectively. NLO and LO results are shown with
solid and dashed lines, respectively.

Figure~\ref{FIG:M} shows the distribution of the invariant mass of the
$\ell^- \ell^+ \gamma$ system $M(\ell^-  \ell^+  \gamma)$. The NLO
corrections are fairly uniform in the invariant mass, so that the NLO
distribution can be approximately described by simply scaling up the
LO distribution by a factor of 1.2 (1.3) at the Tevatron (LHC). The
NLO corrections are larger at the LHC due to the larger gluon density
at the LHC energy.

The rapidity distributions [in the laboratory frame] for the charged
leptons $y(\ell)$ and the $Z$ boson $y(Z)$ are shown in
Figs.~\ref{FIG:YL} and \ref{FIG:YZ}, respectively.  [Both leptons have
been histogrammed in Fig.~\ref{FIG:YL}.] The NLO corrections are
largest in the central rapidity region where the cross section is also
the largest, however, the NLO cross section is uniformly enhanced over
the LO cross section in the entire rapidity range. This fact is
illustrated in Fig.~\ref{FIG:YRATIO} where the ratio of the NLO to LO
cross section is plotted as a function of $y(\ell)$. The ratio has a
constant value of 1.2 (1.3) at the Tevatron (LHC) center of mass
energy.

The angular distributions of the leptonic decay products contain
information on the helicity of the $Z$ boson and are simplest in the
rest frame of the $Z$ boson. Figure~\ref{FIG:COS} shows the polar
angle distribution of the $\ell^-$ in the $Z$ boson rest frame,
measured with respect to the $Z$ boson direction in the $Z\gamma$ rest
frame,  {\it i.e.}, $\cos\theta_{-} = \hat p_{\ell^-} \cdot \hat
p_{Z}^{}$ where  $\hat p_{\ell^-}$ is the unit-normalized
three-momentum  of the $\ell^-$ in the $Z$ boson rest frame and $\hat
p_{Z}^{}$ is the unit-normalized three-momentum  of the $Z$ boson in
the $Z\gamma$ rest frame. If no cuts were imposed on the leptons,
this distribution would exhibit a $1 + \cos^2 \theta_{-}$ dependance,
which is indicative of transversely polarized $Z$ bosons.
Unfortunately, the shapes of the distributions in Fig.~\ref{FIG:COS}
indicate that there is little or no polarization information left in
the distributions. In the presence of cuts, the angular distribution
of leptons produced via the decay of a gauge boson is, in general,
dominated by kinematic effects rather than polarization effects
\cite{MIRKES}. The sharp drops in the distributions near $\cos
\theta_{-} = \pm 1$  are due to the kinematic cuts. The NLO
corrections are once again uniform over the range of $\cos\theta_{-}$.

In the distributions presented so far, the NLO corrections simply
enhanced the LO distributions and produced little or no change in the
shapes of the distributions.  This is not the case for transverse
momentum distributions; instead, the NLO corrections increase with the
transverse momentum. The transverse momentum distributions of the
charged leptons $p_T^{}(\ell)$, the $Z$ boson $p_T^{}(Z)$, and the
photon $p_T^{}(\gamma)$, are shown in Figs.~\ref{FIG:PTL},
\ref{FIG:PTZ}, and \ref{FIG:PTGAMMA}, respectively.  [Both leptons
have been histogrammed in Fig.~\ref{FIG:PTL}.] In all three figures,
the NLO corrections increase as the transverse momentum increases.
The large corrections at high $p_T^{}$ are due to the contributions
from the real emission subprocesses $q \bar q \to Z \gamma g$,
$q g \to Z \gamma q$, and $g \bar q \to Z \gamma \bar q$ which enter
at order $\alpha_s$.

Comparing the curves in Fig.~\ref{FIG:PTL} for the two different
energies, one sees that the curves at the LHC energy do not turn over
at small values of $p_T^{}(\ell)$.  This is simply due to the more
restrictive cuts used for the LHC case.

The NLO and LO $p_T^{}(Z)$ distributions in Fig.~\ref{FIG:PTZ} have
different behaviors at small values of $p_T^{}(Z)$. At LO the $Z$
boson and photon are back-to-back in the plane transverse to the beam
direction, thus the $p_T^{}(Z)$ distribution cuts off at the value of
the minimum $p_T^{}(\gamma)$ cut [$p_T^{}(\gamma) = 10\ (50)$~GeV for
the Tevatron (LHC)].  The NLO cross section on the other hand has a
finite but rapidly decreasing distribution for values of $p_T^{}(Z)$
below the minimum  $p_T^{}(\gamma)$ cut.  In this region of small
$p_T^{}(Z)$, the photon is recoiling against a jet.

In addition to showing the NLO and LO differential cross sections
versus $p_T^{}(\gamma)$,  Fig.~\ref{FIG:PTGAMMA} also shows the 0-jet
and 1-jet components of the NLO inclusive cross section.  Here a jet
is defined as a final state quark or gluon with
\begin{eqnarray}
p_T^{}(j)>10~{\rm GeV}\hskip 1.cm {\rm and} \hskip 1.cm |y(j)|<2.5
\label{EQ:TEVJET}
\end{eqnarray}
at the Tevatron, and
\begin{eqnarray}
p_T^{}(j)>50~{\rm GeV}\hskip 1.cm {\rm and} \hskip 1.cm |y(j)|<3
\label{EQ:LHCJET}
\end{eqnarray}
at the LHC. The sum of the 0-jet and 1-jet cross  sections is equal to
the inclusive NLO cross section.  The decomposition shows that the
1-jet component is small at the Tevatron energy, and is small at low
values of $p_T^{}(\gamma)$ but becomes the dominant component at large
values of  $p_T^{}(\gamma)$ at the LHC energy. Thus most of the NLO
corrections at high $p_T^{}(\gamma)$ are due to events with hard jets.
This is also true for the  $p_T^{}(\ell)$ and  $p_T^{}(Z)$
distributions.

The increase of the NLO corrections with the transverse momentum is
clearly illustrated in Fig.~\ref{FIG:PTRATIO} where the ratio of the
NLO to LO cross section is plotted as a function of  $p_T^{}(\gamma)$.
The ratio increases from 1.2 to 1.5 (1.3 to 2.0) over the range of
$p_T^{}(\gamma)$ shown for the Tevatron (LHC) center of mass energy.

\section{SUMMARY}

The QCD radiative corrections to hadronic $Z\gamma$ production have
been calculated to order $\alpha_s$ with leptonic decays of the $Z$
boson included.  The inclusion of the leptonic decays makes the
calculation more realistic since it is the leptonic decay products
that are observed in an experiment.  Distributions of the final state
decay products have been given for the Tevatron and LHC center of mass
energies.  The calculation includes typical acceptance cuts on the
final state leptons and photon.

The calculation is done by using the Monte Carlo method for NLO
calculations  in combination with helicity amplitude methods.  With
the Monte Carlo method it is easy to impose experimentally motivated
acceptance cuts on the final state particles, also,  it is possible to
calculate the order $\alpha_s$ QCD corrections for exclusive channels,
{\it e.g.},  $p\, p\hskip-7pt\hbox{$^{^{(\!-\!)}}$} \rightarrow Z
\gamma + 0$~jet. The narrow width approximation has been used for the
decaying $Z$ boson.  This simplifies the calculation greatly since it
is possible to ignore contributions from radiative decay diagrams
without violating  electromagnetic gauge invariance.   Futhermore, in
the narrow width approximation it is  particularly easy to extend the
NLO calculation of real  $Z \gamma$ production to include the leptonic
decay of the $Z$ boson.  Spin correlations are included everywhere in
the calculation except in the virtual contribution where they can be
safely neglected.

In general, the  QCD radiative corrections are uniform in invariant
mass and angular distributions; these distributions are scaled up in
magnitude by the corrections, but undergo little change in shape.  In
contrast, the QCD radiative corrections increase with transverse
momentum, and as a result, the $p_T^{}$ distributions are
significantly enhanced at high $p_T^{}$.  The large corrections at
high  $p_T^{}$ are due to contributions from the real emission
processes which enter at order $\alpha_s$. The QCD radiative
corrections increase with the center of mass energy due to the
increasing gluon density in the proton. These behaviors of the NLO
corrections are qualitatively the same as those found in the NLO
corrections to  $W^{\pm} \gamma$ \cite{WGAMMANS},  $ZZ$, $W^- W^+$,
and $W^{\pm}Z$ production \cite{ZZWWWZ}.

%%%%%%%%%%%%%%%%%%%%%%%%% ACKNOWLEDGMENTS %%%%%%%%%%%%%%%%%%%%%%%%%%%%%%%%%%%%
%
\acknowledgements

Discussions with T.~Diehl, S.~Errede, and T.~Muller regarding the
status of $Z\gamma$ production at the Tevatron are gratefully
acknowledged. This work has been supported in part by Department of
Energy grant \#DE-FG03-91ER40674 and by Texas National Research
Laboratory grant \#RGFY93-330.

%\newpage
%
%%%%%%%%%%%%%%%%%%%%%%%%%%%% APPENDIX %%%%%%%%%%%%%%%%%%%%%%%%%%%%%%%%%%%%%%%%%
%
%
%%%%%%%%%%%%%%%%%%%%% REFERENCES %%%%%%%%%%%%%%%%%%%%%%%%%%%%%%%%%%%%%%%%%%%%%%
%

%
%\newpage
%
%%%%%%%%%%%%%%%%%%%%%%%%% TABLES %%%%%%%%%%%%%%%%%%%%%%%%%%%%%%%%%%%%%%%%%%%%
%
\widetext
\begin{table}
\caption{Total LO and NLO cross sections for the process
$p\,p\hskip-7pt\protect{\hbox{$^{^{(\!-\!)}}$}}
\to Z \gamma + X \to \ell^- \ell^+ \gamma + X$,
for center of mass energies corresponding to the present
Tevatron [$\protect{\sqrt{s}=1.8}$~TeV],
an upgraded Tevatron [$\protect{\sqrt{s}=3.5}$~TeV],
and the proposed LHC [$\protect{\sqrt{s}=14}$~TeV].
The cross sections have been
summed over $\ell = e, \mu$ and
the cuts listed in Sec.~IIIB have been imposed.}
\label{TABLE1}
\begin{tabular}{cclc}
 \multicolumn{1}{c}{$\protect{\sqrt{s}}$ (TeV)}
&\multicolumn{1}{c}{$p\,p\hskip-7pt\protect{\hbox{$^{^{(\!-\!)}}$}}$}
&\multicolumn{1}{c}{ }
&\multicolumn{1}{c}{$\sigma$ (fb)}\\
\tableline
1.8 & $p\, \bar p$ & LO  & 360 \\
    &              & NLO & 420 \\
\tableline
3.5 & $p\, \bar p$ & LO  & 540 \\
    &              & NLO & 620 \\
\tableline
14. & $p\, p$      & LO  & 230 \\
    &              & NLO & 310 \\
\end{tabular}
\end{table}
%
%\newpage
%
%%%%%%%%%%%%%%%%%%%%%% FIGURE CAPTIONS %%%%%%%%%%%%%%%%%%%%%%%%%%%%%%%%%%%%%%
%
\begin{figure}
\caption{Invariant mass distribution of the $\ell^- \ell^+ \gamma$
system for the process
$p\,p\hskip-7pt\hbox{$^{^{(\!-\!)}}$} \rightarrow
Z \gamma + X \rightarrow \ell^- \ell^+ \gamma + X$.
Parts a) and b) are for the Tevatron and LHC center of mass energies,
respectively.  The NLO (solid line) and LO (dashed line) cross sections
are shown.  The cross sections have been summed over $\ell = e, \mu$
and the cuts listed in Sec.~IIIB have been imposed.}
\label{FIG:M}
\end{figure}

\begin{figure}
\caption{Same as Fig.~\protect{\ref{FIG:M}} but for the
rapidity distribution of the leptons.}
\label{FIG:YL}
\end{figure}

\begin{figure}
\caption{Same as Fig.~\protect{\ref{FIG:M}} but for the
rapidity distribution of the $Z$ boson.}
\label{FIG:YZ}
\end{figure}

\begin{figure}
\caption{The ratio
$[d\sigma^{\hbox{\protect{\scriptsize NLO}}} / dy(\ell)] /
 [d\sigma^{\hbox{\protect{\scriptsize  LO}}} / dy(\ell)]$
plotted versus $y(\ell)$ for the process
$p\,p\hskip-7pt\hbox{$^{^{(\!-\!)}}$} \rightarrow
Z \gamma + X \rightarrow \ell^- \ell^+ \gamma + X$.
Parts a) and b) are for the Tevatron and LHC center of mass energies,
respectively. The cuts listed in Sec.~IIIB have been imposed.}
\label{FIG:YRATIO}
\end{figure}

\begin{figure}
\caption{Same as Fig.~\protect{\ref{FIG:M}} but for the
angular distribution of the negatively charged lepton.
The angle $\theta_{-}$ is measured in the $Z$ boson rest frame with
respect to the $Z$ boson direction in the $Z\gamma$ rest frame.}
\label{FIG:COS}
\end{figure}

\begin{figure}
\caption{Same as Fig.~\protect{\ref{FIG:M}} but for the
transverse momentum distribution of the charged leptons.}
\label{FIG:PTL}
\end{figure}

\begin{figure}
\caption{Same as Fig.~\protect{\ref{FIG:M}} but for the
transverse momentum distribution of the $Z$ boson.}
\label{FIG:PTZ}
\end{figure}

\begin{figure}
\caption{Same as Fig.~\protect{\ref{FIG:M}} but for the
transverse momentum distribution of the photon.
In addition, the 0-jet (dotted line) and 1-jet (dot-dashed line)
cross sections are also shown.}
\label{FIG:PTGAMMA}
\end{figure}

\begin{figure}
\caption{The ratio
$[d\sigma^{\hbox{\protect{\scriptsize NLO}}} / dp_T^{}(\gamma)] /
 [d\sigma^{\hbox{\protect{\scriptsize  LO}}} / dp_T^{}(\gamma)]$
plotted versus $p_T^{}(\gamma)$ for the process
$p\,p\hskip-7pt\hbox{$^{^{(\!-\!)}}$} \rightarrow
Z \gamma + X \rightarrow \ell^- \ell^+ \gamma + X$.
Parts a) and b) are for the Tevatron and LHC center of mass energies,
respectively. The cuts listed in Sec.~IIIB have been imposed.}
\label{FIG:PTRATIO}
\end{figure}

%
%%%%%%%%%%%%%%%%%%%%%%%%%%%%%%%%%%%%%%%%%%%%%%%%%%%%%%%%%%%%%%%%%%%%%%%%%%%%%
%

\begin{references}

\bibitem{EHLQ}
E.~Eichten, I.~Hinchliffe, K.~Lane, and C.~Quigg,
Rev. Mod. Phys. {\bf 56}, 579 (1984); {\bf 58}, 1065(E) (1986).  % general

\bibitem{LEURER}
M.~Leurer, H.~Harari, and R.~Barbieri,
Phys. Lett. B {\bf 141}, 455 (1985). 	% composite Z

\bibitem{ANOMALOUS}
U.~Baur and E.~L.~Berger,
Phys. Rev. D {\bf 47}, 4889 (1993);
Z.~Ryzak, Nucl. Phys. {\bf B289}, 301 (1987);
H.~Baer, V.~Barger, and K.~Hagiwara,
Phys. Rev. D {\bf 30}, 1513 (1984).  	% anomalous Z gamma

\bibitem{RENARD}
F.~Renard,
Nucl. Phys. {\bf B196}, 93 (1982).  	% Z gamma

\bibitem{BREM}
J.~Ohnemus and W.~J.~Stirling,
Phys. Lett. B {\bf 298}, 230 (1993). 	% photon brem

\bibitem{VVONEJET}
U.~Baur, E.~W.~N.~Glover, and J.~J. van der Bij,
Nucl. Phys. {\bf B318}, 106 (1989). % VV + 1 jets

\bibitem{VVTWOJET}
V.~Barger, T.~Han, J.~Ohnemus, and D.~Zeppenfeld,
Phys. Rev. D {\bf 41}, 2782 (1990).  % VV + 2 jets

\bibitem{AGPT}
Ll.~Ametller, E.~Gava, N.~Paver, and D.~Treleani,
Phys. Rev. D {\bf 32}, 1699 (1985). % gg -> Z gamma

\bibitem{GLUONFUSION}
J.~J.~van der Bij and E.~W.~N.~Glover,
Phys. Lett. B {\bf 206}, 701 (1988). % gg -> Z gamma

\bibitem{ZGAMMA}
J.~Ohnemus,
Phys. Rev. D {\bf 47}, 940 (1993). 	% NLO W gamma and Z gamma

\bibitem{ERREDE}
S.~Errede and T.~Muller, private communication.

\bibitem{DIEHL}
T.~Diehl, private communication.

\bibitem{NLOMC}
J.~Ohnemus and J.~F.~Owens,
Phys. Rev. D {\bf 43}, 3626 (1991);	% NLO ZZ
J.~Ohnemus,
Phys. Rev. D {\bf 44}, 1403 (1991);	% NLO WW
J.~Ohnemus,
Phys. Rev. D {\bf 44}, 3477 (1991); % NLO WZ
H.~Baer, J.~Ohnemus, and J.~F.~Owens,
Phys. Rev. D {\bf 40}, 2844 (1989);  	% photoproduction
H.~Baer, J.~Ohnemus, and J.~F.~Owens,
Phys. Rev. D {\bf 42}, 61 (1990);
Phys. Lett. B {\bf 234}, 127 (1990);	% directphoton
H.~Baer and M.~H.~Reno,
Phys. Rev. D {\bf 43}, 2892 (1991);	% NLO W
B.~Bailey, J.~Ohnemus, and J.~F.~Owens,
Phys. Rev. D {\bf 46}, 2018 (1992);	% NLO gamma gamma
J.~Ohnemus and W.~J.~Stirling,
Phys. Rev. D {\bf 47}, 2722 (1993);  	% NLO WH
H.~Baer, B.~Bailey, and J.~F.~Owens,
Phys. Rev. D {\bf 47}, 2730 (1993);	% NLO WH
L.~Bergmann,
Ph.D. dissertation, Florida State University,
report No. FSU-HEP-890215, 1989 (unpublished).

\bibitem{HELICITY}
P.~De Causmaecker, R.~Gastmans, W.~Troost, and T.~T.~Wu,
Phys. Lett. B {\bf 105}, 215 (1981); Nucl. Phys. {\bf B206}, 53 (1982);
F.~A.~Berends, R.~Kleiss, P.~De Causmaecker, R.~Gastmans, W.~Troost,
and T.~T.~Wu, Nucl. Phys. {\bf B206}, 61 (1982);
CALKUL Collaboration, F.~A.~Berends {\it et al.},
Nucl. Phys. {\bf B239}, 382 (1984);
K.~Hagiwara and D.~Zeppenfeld, Nucl. Phys. {\bf B274}, 1 (1986);
Z.~Xu, D.-H.~Zhang and L.~Chang, Nucl. Phys. {\bf B291}, 392 (1987);
R.~Gastmans and T.~T.~Wu, {\it The Ubiquitous Photon: Helicity Method
for QED and QCD} (Oxford University Press, Oxford, 1990).

\bibitem{WGAMMANS}
U.~Baur, T.~Han, and J.~Ohnemus,
Phys. Rev. D {\bf 48}, 5140 (1993).	% W gamma non-standard

\bibitem{TOPQUARK}
CDF Collaboration, F.~Abe {\it et al.},
FERMILAB-PUB-94/116-E; FERMILAB-PUB-94/097-E.

\bibitem{LEP}
LEP Collaborations: ALEPH, DELPHI, L3, OPAL, and the LEP Electroweak
Working Group, {\it Updated Parameters of the $Z^0$ Resonance from
Combined Preliminary Data of the LEP Experiments},
CERN/PPE/93-157; Phys. Lett. B {\bf 276}, 247 (1992).

\bibitem{SLD}
SLD Collaboration, K.~Abe {\it et al.}, SLAC-PUB-6456.

\bibitem{SPPS}
UA2 Collaboration, J.~Alitti {\it et al.},
Phys. Lett. B {\bf 241}, 150 (1990);
UA1 Collaboration, G.~Arnison {\it et al.},
Europhys. Lett. {\bf 1}, 327 (1986).

\bibitem{TEVATRON}
CDF Collaboration, F.~Abe {\it et~al.},
Phys. Rev. Lett. {\bf 65}, 2243 (1990);
Phys. Rev. D {\bf 43}, 2070 (1991);
D0 Collaboration, N.~A.~Graf, FERMILAB-CONF-94-021-E;
Q.~Zhu, FERMILAB-CONF-93-396-E.

\bibitem{MRSA}
A.~D.~Martin, R.~G.~Roberts, and W.~J.~Stirling,
RAL-94-055, DTP/94/34.			% MRS prime

\bibitem{MSBAR}
W.~A.~Bardeen, A.~J.~Buras, D.~W.~Duke, and T.~Muta,
Phys. Rev. D {\bf 18}, 3998 (1978).  	% MS bar scheme

\bibitem{ISOLATED1}
CDF Collaboration, F.~Abe {\it et~al.},
Phys. Rev. Lett. {\bf 68}, 2734 (1992);
Phys. Rev. D {\bf 48}, 2998 (1993).	% photon isolation cut

\bibitem{ISOLATED2}
P.~Aurenche, R.~Baier, and M.~Fontannaz,
Phys. Rev. D {\bf 42}, 1440 (1990);
P.~Aurenche {\it et~al.}, in {\it Proceedings of the
Large Hadron Collider Workshop}, Aachen, Germany, 1990, edited by
G.~Jarlskog and D.~Rein, CERN 90-10, Vol.~II, p.~69.	% photon isolation cut

\bibitem{MIRKES}
E.~Mirkes and J.~Ohnemus, MAD/PH/834, UCD--94--23.

\bibitem{ZZWWWZ}
J.~Ohnemus, to appear in Phys. Rev. D.

\end{references}
\end{document}